\documentclass[conference]{IEEEtran}
\ifCLASSINFOpdf
   \usepackage[pdftex]{graphicx}
\else
   \usepackage[dvips]{graphicx}
\fi

\usepackage{amsmath}

\usepackage[tight,footnotesize]{subfigure}

\usepackage{fancyhdr}
\renewcommand{\thispagestyle}[1]{} 

\begin{document}

%
\title{\LARGE{ELECTROMAGNETIC WAVE PROPAGATION IN THE PLASMA LAYER OF A REENTRY VEHICLE}}

\author{\IEEEauthorblockN{M. Kundrapu\IEEEauthorrefmark{2}, J. Loverich, K. Beckwith, P. Stoltz}
\IEEEauthorblockA{\textit{Tech-X Corporation, 5621 Arapahoe Ave.}\\
\textit{Boulder, CO, USA}}

\IEEEauthorblockN{A. Shashurin, M. Keidar}
\IEEEauthorblockA{\textit{The George Washington University, 801 22nd Street, NW}\\
\textit{Washington, DC, USA}}
}


%


\maketitle

\begin{abstract}
The ability to simulate a reentry vehicle plasma layer and the radio wave interaction with that layer, is crucial to the design of aerospace vehicles when the analysis of radio communication blackout is required. Results of aerothermal heating, plasma generation and electromagnetic wave propagation over a reentry vehicle are presented in this paper. Simulation of a magnetic window radio communication blackout mitigation method is successfully demonstrated.
\end{abstract}


%
\IEEEpeerreviewmaketitle

\footnotetext[2]{Email: madhusnk@txcorp.com}

\section{Introduction}
During reentry into the atmosphere, aerospace vehicles are subjected to severe aerothermal heating which causes the surrounding gas to ionize and form a plasma. The analysis of the plasma around the vehicle is useful in developing new strategies and mitigation devices to minimize blackout related issues.\cite{meyer2007system,gillman2010review} Modeling aspects and the simulation results of aerothermal heating, plasma generation and electromagnetic (EM) wave propagation over a RAM-C reentry vehicle are presented. A magnetic window blackout-mitigation scheme is demonstrated.\cite{thoma2009electromagnetic,white1974amplification,stenzel2013new}

In this paper, the partially ionized air plasma distribution is generated using a single fluid, multi-species model, with average thermophysical properties. Navier-Stokes equations are solved for the flow mixture. The individual species mass conservation equations are then solved to track their densities. The species considered in the mixture are $N_2$, $O_2$, $N$, $O$, $NO$, $NO^+$, and $e$ (electrons).\cite{josyula2003governing} Chemical reaction rate equations are solved to account for the variations in densities of the species.  After the plasma distribution is determined, Maxwell's equations combined with a non-neutral multi-fluid model are solved for the propagation of electro-magnetic wave in the plasma.  A finite volume method is used to solve the conservation equations.   The interface fluxes are obtained using HLLE scheme. The variables are interpolated on the the cell face using a MUSCL scheme and the possible spurious oscillations are limited using Van-Leer limiter. A TVD Runge-Kutta third order scheme is used for time integration.  The problem involves three time scales 1) fluid convection-diffusion, 2) chemical reactions, and 3) EM wave propagation and plasma oscillation time scale.  The Navier-Stokes equations are solved with the time step restricted by CFL condition and viscous-thermal diffusion. The rate equations are integrated separately on a smaller time scale owing to the reactions. Maxwells equations are solved on a further smaller time scale when the plasma density distribution has reached quasi-steady state. Though all of the equations can be solved simultaneously at the smallest time scale, the trade-off between computational time and accuracy of solution demands splitting the simulation. For instance, the fluid time scale is five orders of magnitude larger than that of EM wave propagation time scale. The simulations were performed using the fluid-plasma multiphysics simulation tool, USim.\cite{loverich2013nautilus, kundrapu2013modeling, shashurin2014laboratory}

\hfill

\section{Blackout simulation models}
\subsection{Multi-species single fluid model}
The hypersonic flow is simulated using the multi-species single fluid transport equations given by Eqs.(\ref{eq:mass})--(\ref{eq:speciesCont}). Where, $\rho$, $\vec{u}$, $p$, and $e$ are the mass density, velocity vector, pressure, and the total energy density respectively. The total energy is the sum of internal energy, kinetic energy, and the species chemical energy. In the RHS of Eq.(\ref{eq:momentum}) $\tau$ is the shear stress tensor.  
\begin{equation}
\frac{\partial \rho}{\partial t} + \nabla\cdot\left(\rho \vec{u}\right) = 0
\label{eq:mass}
\end{equation}
\begin{equation}
\frac{\partial \left(\rho \vec{u}\right)}{\partial t} + \nabla\cdot\left(\rho \vec{u}\vec{u} + pI\right) = \nabla\cdot\mathbf{\tau}
\label{eq:momentum}
\end{equation}
\begin{equation}
\frac{\partial \left(e\right)}{\partial t} + \nabla\cdot\left(\vec{u}\left(e+p\right)\right) = \nabla\left(\tau \cdot \vec{u}\right) + \nabla\cdot\left(k\nabla T\right)
\label{eq:totalEnergy}
\end{equation}
\begin{equation}
\frac{\partial n_i}{\partial t} + \nabla\cdot\left(\vec{u} n_i\right) = s_i
\label{eq:speciesCont}
\end{equation}

where,\ $\rho = \sum\limits_{i}{n_{i}m_{i}}$, $p = \rho R T$, $\tau =  -\frac{2}{3}\mu\left(\nabla\cdot\vec{u}\right)I + \mu\left(\nabla\vec{u} + \left(\nabla\vec{u}\right)^{T} \right)$ and \\ $e = \frac{p}{\gamma -1} + \frac{1}{2}\rho \vec{u}\cdot\vec{u} +\sum\limits_{i}{n_{i}H_{i}}$.

The mass conservation of the individual species in the bulk fluid is satisfied separately for each of the species using Eq. (\ref{eq:speciesCont}). The velocity $\vec{u}$ is the same as that of the bulk fluid. The right hand side of Eq. (\ref{eq:speciesCont}) represents the rate of change of species density due to the chemical reactions.

The properties, viscosity $\mu$, thermal conductivity $k$ and the specific heat $c_p$ of the individual species were obtained from the kinetic theory of gases as given by the Eqs. (\ref{eq:viscosity})--(\ref{eq:specificHeat}). The fluid thermal conductivity and viscosity in Eqs. (\ref{eq:mass})--(\ref{eq:totalEnergy}) were obtained using mole fraction averaging. In Eq.(\ref{eq:viscosity}), $i$ represents the index of species, $m$ is the mass of the particle, $k_B$ is the Boltzmann constant, $T$ is the temperature, $\sigma$ is the collision diameter, and $\Omega$ is the collision integral. The hard sphere collision diameter is used in this work.  The specific heat $c_p$ was obtained using the mass fraction averaging. The gas constant $R$ was computed using the mole fraction averaged molecular weight and the universal gas constant. 

\begin{equation}
\mu_{i} = \frac{5}{16}\frac{\sqrt{\pi m_{i}k_{B}T}}{\left(\pi \sigma^2 \Omega\right)}
\label{eq:viscosity}
\end{equation}
\begin{equation}
k_{i} = \frac{5}{2}{c_{v}}_{i}\mu_{i}
\label{eq:thermalConductivity}
\end{equation}
\begin{equation}
{c_{p}}_{i} = \left(\frac{f}{2}+1\right)R_{i}
\label{eq:specificHeat}
\end{equation}

\begin{equation}
{c_{v}}_{i} = {c_{p}}_{i}-R_{i}
\label{eq:specificHeat}
\end{equation}

\subsection{Multi-fluid transport}
EM wave propagation was carried out using multi-fluid transport and Maxwell's equations.\cite{shumlak2003approximate,hakim2006high} The fluid transport equation are shown given by Eqs. (\ref{eq:twoFluidMass})--(\ref{eq:twoFluidTotalEnergy}).\cite{zhdanov2002transport,shashurin2014laboratory}. The index $\alpha$ is for any fluid. In Eq.\ref{eq:twoFluidMomentum}, $q$ is the unit electric charge, $\vec{E}$ is the electric field vector, and $\vec{B}$ is the magnetic field vector. The fluids are coupled using momentum, kinetic energy, and internal energy exchange terms given by Eqs. (\ref{eq:momExch})--(\ref{eq:intExch}) respectively. The index $i$ is for any species other than $\alpha$. $\mu_{\alpha i}$ and $\zeta_{\alpha i}$ are the reduced mass and the collision time between species $\alpha$ and $i$ respectively. The electric and magnetic fields are solved using Maxwell's equations Eq.\ref{eq:ampere} and Eq.\ref{eq:faraday} along with the electric and magnetic field divergences Eq.(\ref{eq:electricDivergence}), Eq. (\ref{eq:magneticDivergence}).
\begin{equation}
\frac{\partial \rho_{\alpha}}{\partial t} + \nabla\cdot\left(\rho_{\alpha} \vec{u}_{\alpha}\right) = 0
\label{eq:twoFluidMass}
\end{equation}

\begin{equation}
\begin{split}
\frac{\partial \left(\rho_{\alpha} \vec{u}_{\alpha}\right)}{\partial t} + \nabla\cdot\left(\rho_{\alpha} \vec{u}_{\alpha}\vec{u}_{\alpha} + p_{\alpha}I\right)& = \frac{\rho_{\alpha}}{m_{\alpha}} q_{\alpha} \left(\vec{E} +\vec{u}_{\alpha}\times\vec{B} \right) \\
& \quad + \nabla\cdot\mathbf{\tau}_{\alpha} + \vec{R}_{\alpha}
\end{split}
\label{eq:twoFluidMomentum}
\end{equation}

\begin{equation}
\begin{split}
\frac{\partial \left(e_{\alpha}\right)}{\partial t} + \nabla\cdot\left(\vec{u}_{\alpha}\left(e+p_{\alpha}\right)\right)& = \frac{\rho_{\alpha}}{m_{\alpha}} q_{\alpha}\vec{u}_{\alpha}\cdot\vec{E} \\
& \quad + \nabla\left(\tau_{\alpha}\cdot\vec{u}_{\alpha}\right) + \nabla\cdot\left(k_{\alpha}\nabla T_{\alpha}\right) \\
& \quad + \vec{V}_{\alpha}\cdot\vec{R}_{\alpha} + Q_{\alpha}
\end{split}
\label{eq:twoFluidTotalEnergy}
\end{equation}
where,
\begin{equation}
\vec{V}_{\alpha} = \left(\sum\limits_{i}\rho_{i}\vec{u}_{i}\right)/\sum\limits_{i}\rho_{i}
\label{eq:momExch}
\end{equation}
\begin{equation}
\vec{R}_{\alpha} = - \sum\limits_{i}\frac{\rho_{\alpha}}{m_{\alpha}}\mu_{\alpha i}\zeta_{\alpha i}^{-1}\left(\vec{u}_{\alpha}-\vec{u}_{i}\right)
\label{eq:kinExch}
\end{equation}
and
\begin{equation}
Q_{\alpha} = - \sum\limits_{i}3k_{B}\frac{\rho_{\alpha}}{m_{\alpha}}\left[\mu_{\alpha i}/\left(m_{\alpha}+m_{i}\right)\right]\zeta_{\alpha i}^{-1}\left(T_{\alpha}-T_{i}\right)
\label{eq:intExch}
\end{equation}

\begin{equation}
\frac{\partial \vec{E}}{\partial t}-c^{2}\nabla\times\vec{B} = -\frac{1}{\epsilon_{0}}\sum\limits_{\alpha}\frac{q_{\alpha}\rho_{\alpha}\vec{u}_{\alpha}}{m_{\alpha}}
\label{eq:ampere}
\end{equation}

\begin{equation}
\frac{\partial \vec{B}}{\partial t}+\nabla\times\vec{E} = 0
\label{eq:faraday}
\end{equation}

\begin{equation}
\nabla\cdot\vec{E}= \frac{1}{\epsilon_{0}}\sum\limits_{\alpha}\frac{q_{\alpha}\rho_{\alpha}}{m_{\alpha}}
\label{eq:electricDivergence}
\end{equation}

\begin{equation}
\nabla\cdot\vec{B}= 0
\label{eq:magneticDivergence}
\end{equation}

\section{Results and Discussion}

\begin{figure}
\includegraphics[width=0.45\textwidth]{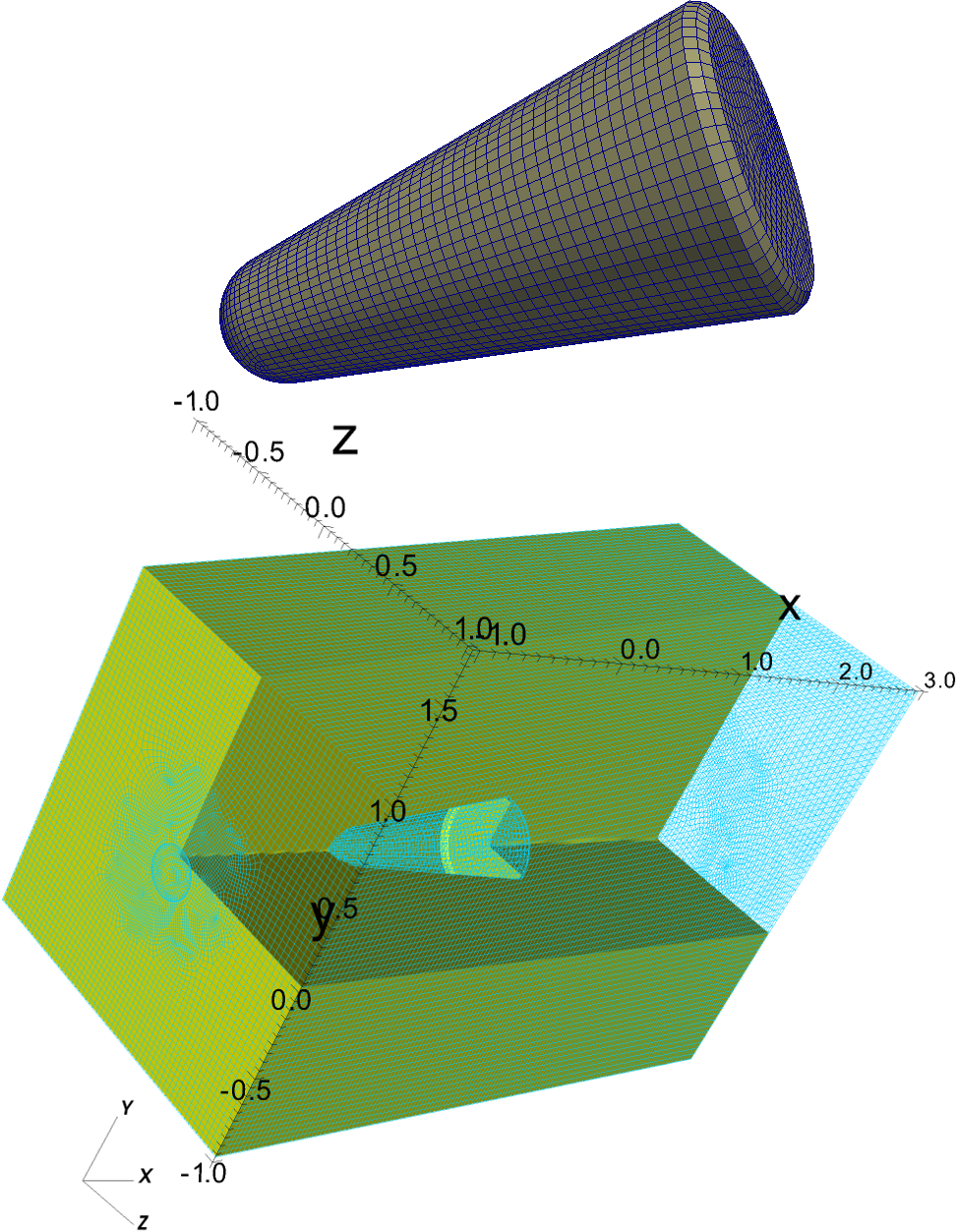}
\caption[]{The unstructured grid used for simulating the hypersonic flow and communication blackout. The grid has 853000 hexahedral cells with the average minimum edge of 7 mm and average maximum edge of 4 cm. The upper figure shows the surface grid of RAM-C.}
\label{fig:grid3d}
\end{figure}

Figure \ref{fig:grid3d} shows the grid used to simulate the multi-species and EM wave propagation over the RAM-C.  The grid was generated using Cubit meshing software. The grid has 853,000 hexahedral cells with an average minimum edge of 7 mm and an average maximum edge of 4 cm. The upper figure shows the surface grid of the RAM-C. The total length of RAM-C is 1.295 m with 81$^\circ$ cone of spherical nose cap of radius 0.1524 m and a base radius of 0.335 m. The overall domain is 4 m long with 2$\times$2 $m^2$ front and 2$\times$2.75 $m^2$ cross section.  

\begin{figure}
\includegraphics[width=0.45\textwidth]{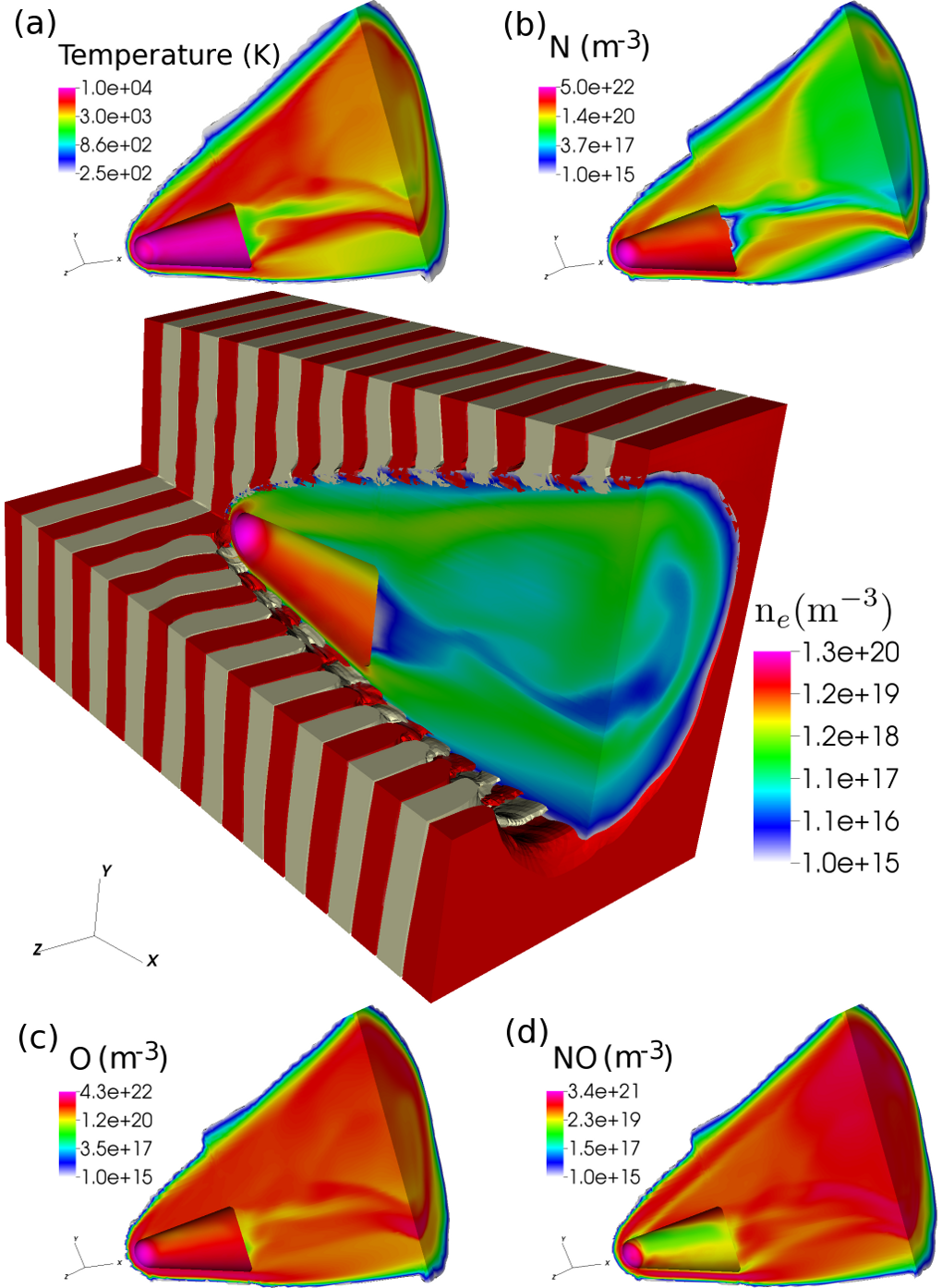}
\caption[]{The electromagnetic wave reflection from the over dense plasma layer of RAM C reentry vehicle. The main plot shows the plasma layer density (electrons) along with the $E_x$ of the EM wave. The subplots (a) temperature distribution, (b) nitrogen atoms (c) oxygen atoms, and (d) nitric oxide. }
\label{fig:3d}
\end{figure}

The multi-species simulation of the RAM-C at an angle of attack of 15$^{\circ}$ is shown in the Fig. \ref{fig:3d}. The RAM-C was flying at an altitude of 61 km at Mach 23. The boundary conditions are standard, no-slip on the wall, inflow at the inlet and out flow on the remaining boundaries. A plane sinusoidal EM wave with frequency $f$=0.8 GHz was excited at the front boundary face of the domain (refer Fig.\ref{fig:grid3d}). The components of electric and magnetic fields at the boundary E$_y$ = $c a_{0}sin(2\pi f t)$, B$_z$ = E$_y$/c. The remaining components are zero. Conductor boundary condition is used on the wall for the Maxwell's equations and copy boundary condition everywhere else. Figure 1 shows the plasma distribution and EM wave propagation around RAM-C.  The peak density of plasma is observed near the stagnation region of nose cap, where highest temperatures are observed as well. The peak density of the plasma is 1.3$\times 10^{20}$ m$^{-3}$. The peak temperature of gas is 10470 K as shown in the subplot (a). The dissociated nitrogen and oxygen are shown in subplots (b) and (c). Their peak values are 5$\times 10^{22}$ and 4.3$\times 10^{22}$ m$^{-3}$ respectively. Note that the dissociation fraction of oxygen is greater than that of nitrogen. The nitric oxide density distribution is shown in subplot (d). The E$_y$ of the EM wave shown in the figure with red and yellow contours correspond to positive and negative amplitudes propagates uninterrupted until it reaches the plasma layer of the RAM-C. The wave is then reflected by the plasma as shown in the XY and XZ cut planes. The EM wave is completely reflected by the plasma as the density grows beyond 8$\times 10^{15}$ m$^{-3}$. The amplitudes values of the wave in Fig.\ref{fig:3d} were limited to $\pm$15 for a better representation of the wave reflection from the plasma layer. Otherwise, it is quite confusing to visualize the waveform in the contours because of the mixing of reflected components E$_x$, E$_z$ with the the source component E$_y$.  In addition, the wave's amplitude increases by about 10 times at the edge of the plasma layer due to the resonance of the evanescent wave.\cite{white1974amplification} The amplified wave propagates along the plasma layer's edge. However, the limited amplitude depiction in Fig.\ref{fig:3d} shows no traces of the wave on the vehicle's surface. Another important issue in the simulation is that, the grid shown in Fig. \ref{fig:grid3d} is under-resolved as our computational resources were limited. Due to the coarseness, the shock wave, with a standoff distance of about 2 cm, is not properly resolved. We expect better results with a resolved grid. Overall, the simulation provides a good understanding of the flow, plasma density distribution and its approximate peak value. Also, the simulation shows that the radio wave does not penetrate to the vehicle surface, even in the wake region. There is no open path, connecting the vehicle surface and the region of the plasma layer along which the plasma frequency is less than the wave frequency. Consequently, one should use active or passive mitigation techniques to mitigate radio blackout in this case.
 
\begin{figure}
\begin{center}
\subfigure[]{
\includegraphics[width=0.215\textwidth]{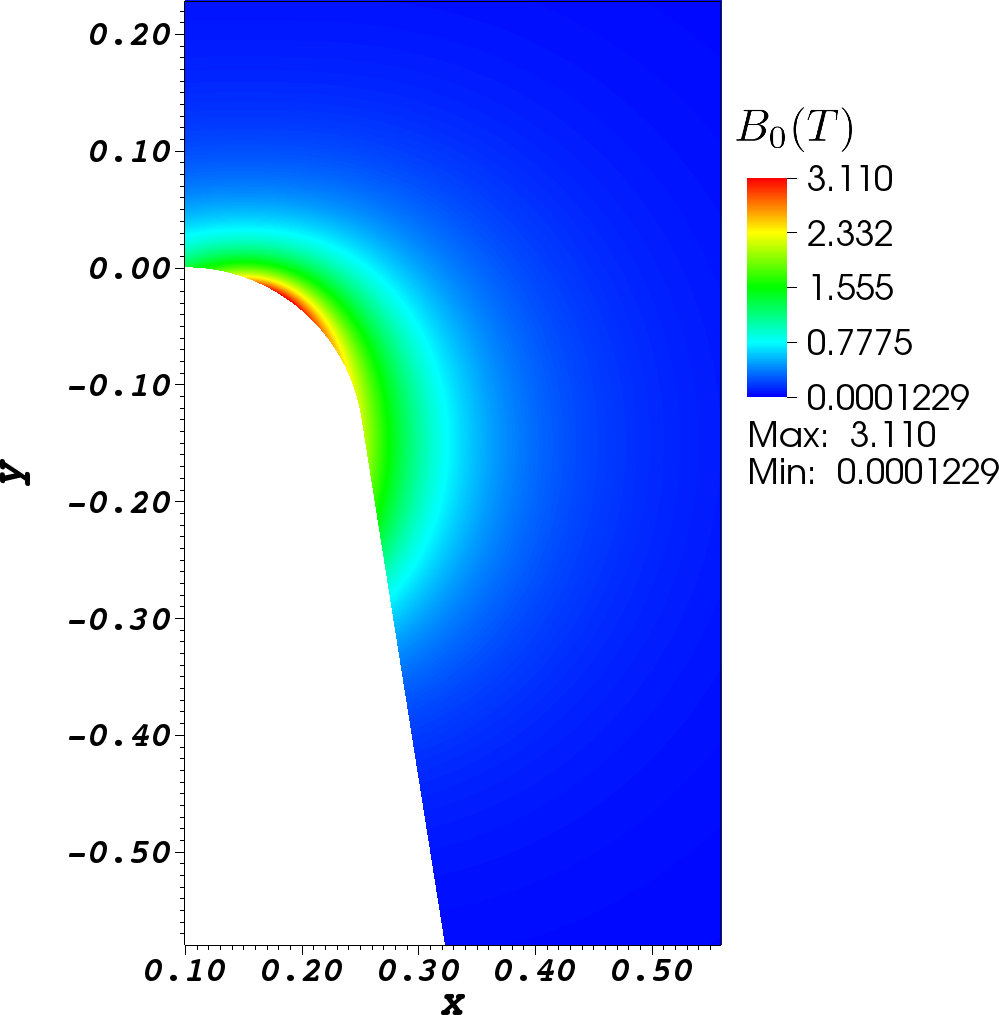}
\label{fig:magB0}}
\quad
\subfigure[]{
\includegraphics[width=0.215\textwidth]{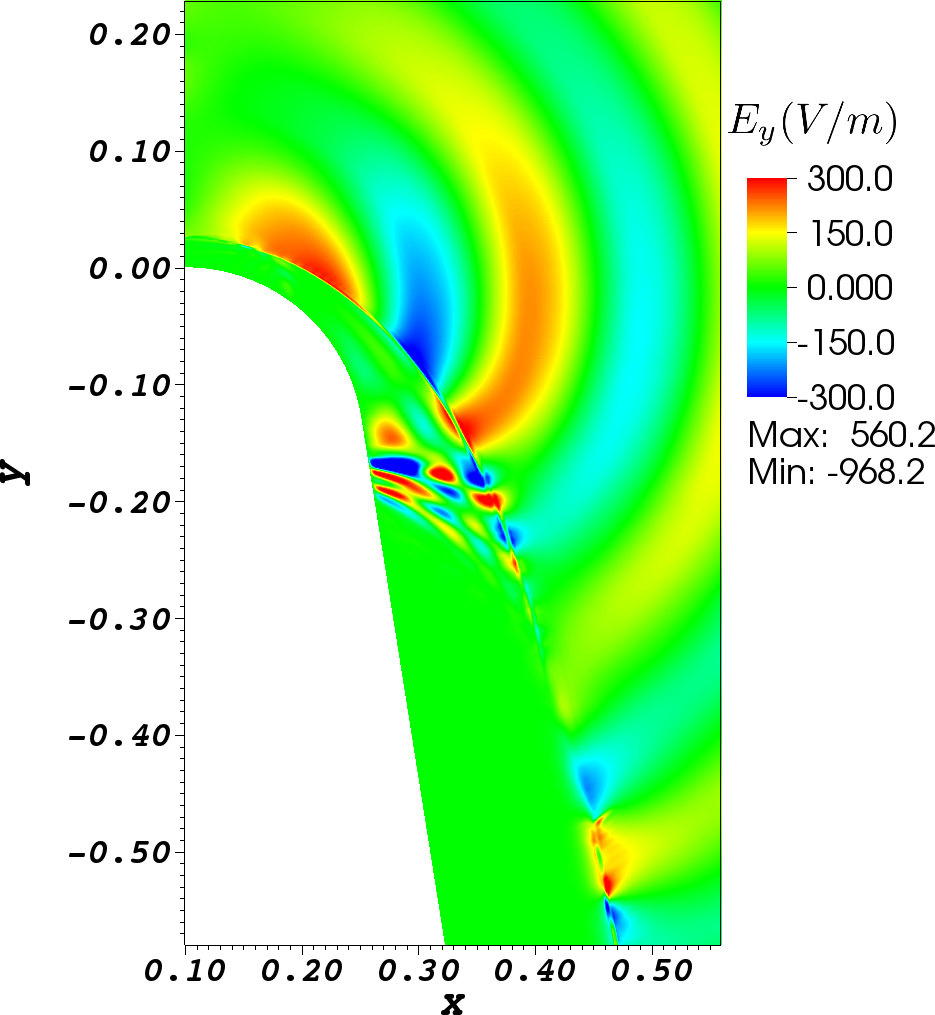}
\label{fig:magEy}}
\caption[]{The \subref{fig:magB0} background static magnetic field and	\subref{fig:magEy} the y (reflected) component of EM wave.}
\vspace{-0.2in}
\label{fig:em2d}
\end{center}
\end{figure}

In order for the wave to propagate on to the vehicle's surface, the magnetic window or whistler wave technique is used. Whistler waves propagate in high dense plasmas along the magnetic field lines. Fig. \ref{fig:em2d} shows the setup of the magnetic window that can allow the EM wave to pass to the surface of the vehicle. The results of a 2d simulation are presented here as we were able to achieve sufficiently high resolution. The wave frequency was doubled to keep it close to the practical frequencies used in real hypersonics operations. A static magnetic field was imposed near the nose cap using a current carrying coil placed inside the nosecap. The coil radius was 0.1 m placed at the geometrical location (0.15m, -0.15m). The current inside the coil was $10^6$ A. As shown in Fig.\ref{fig:magB0}, the coil generates a magnetic field of 3 T on the surface which drops down to 0.8 T at the edge of the plasma layer, where the wave reflects originally.  This field strength is beyond the reach of permanent magnets, however, at lower Mach numbers where the plasma density is lower (as would be the case for many hypersonic vehicles), a much weaker field would be required. Figure \ref{fig:em2d} shows the whistler mode propagation of the wave on to the surface. The y component of the electric field, $E_y$, is shown here. The wave's quality on the surface can be assessed by recording the wave history and computing the frequency using Fourier transforms. The frequency match with the original wave is good indicator that the signal is not too noisy to use.  In the present configuration, estimates show that the signal frequency at the surface matches that of the source signal.

\section{Conclusion}
Detailed three dimensional simulations were performed to obtain the plasma density distribution around the RAM-C reentry vehicle. The simulations were performed for flight conditions at 61 km altitude, Mach 23 and angle of attack $15^\circ$. The plasma distribution simulations were performed on a fully unstructured grid using a single fluid multi-species solver of USim. The EM wave reflection from the edge of the free plasma layer was demonstrated using the calculated plasma distribution to initialize a multi-fluid EM solver. The analysis shows that the radio signal is completely blocked when no mitigation technique is used.  One approach to solving this problem is the use of a magnetic window.  This approach is demonstrated in a simulation which shows that the signal wave propagates on to the vehicle's surface in the whistler mode.






%

\end{document}